# Variable selection for transportability

Megha L. Mehrotra, M. Maria Glymour, Elvin Geng, Daniel Westreich, David V. Glidden

**Correspondence to**:
Megha Mehrotra, PhD
mmehrotra@berkeley.edu
805-501-9056

**Conflicts of interest and sources of funding**: This work was supported by grant 5F31 MH111346 to investigator MLM from the National Institute of Mental Health. The iPrEx study (NCT00458393) was supported by the US National Institute of Health (AI064002 to RMG) and the Bill & Melinda Gates Foundation. This work was also supported by grants *DP2* HD084070 to DW, K24 AI134413 to EG, CFAR SWG Implementation Science Working Group to EG, and AI 126597 to DVG from The National Institutes of Health.

ABSTRACT

       Transportability provides a principled framework to address the problem of applying study results to new populations. Here, we consider the problem of selecting variables to include in transport estimators. We provide a brief overview of the transportability framework and illustrate that while selection diagrams are a vital first step in variable selection, these graphs alone identify a sufficient but not strictly necessary set of variables for generating an unbiased transport estimate. Next, we conduct a simulation experiment assessing the impact of including unnecessary variables on the performance of the parametric g-computation transport estimator. Our results highlight that the types of variables included can affect the bias, variance, and mean squared error of the estimates. We find that addition of variables that are not causes of the outcome but whose distributions differ between the source and target populations can increase the variance and mean squared error of the transported estimates. On the other hand, inclusion of variables that are causes of the outcome—regardless of whether they modify the causal contrast of interest or differ in distribution between the populations—reduces the variance of the estimates without increasing the bias. Finally, exclusion of variables that cause the outcome but do not modify the causal contrast of interest does not increase bias. These findings suggest that variable selection approaches for transport should prioritize identifying and including all causes of the outcome in the study population rather than focusing on variables whose distribution may differ between the study sample and target population.

The transportability framework, which builds on the theoretical foundations of causal inference, outlines the necessary rules and assumptions for determining when and how a causal effect estimated in a study population can be applied to an external target population, that is any population that is distinct from the population that gave rise to the original study sample.[1] These same tools can also be used to better understand observed effect heterogeneity within a study.[2-4]

Ideally, subject matter expertise and a clear understanding of the study and target populations would be the primary guide for variable selection decisions. Selection diagrams—the causal graphs used for transport-- encode prior knowledge of the underlying causal mechanisms and how the study and a target population might differ. Once a selection diagram is drawn, d-separation rules are applied to identify a set of variables that is *sufficient* to transport an effect from a study population to a given target population. However, because of uncertainty about the underlying causal structure or mechanisms in real-world applications, selection diagrams alone may be insufficient to narrow down the list of candidate variables to include only those that are *necessary* for a given application. This is critical because the inclusion of unnecessary variables might affect the performance of transport estimators, but to our knowledge, little has been written about alternative variable selection strategies for transport.

Here, we provide a practical guide to variable selection for transportability. We begin by briefly reviewing the transport framework and graphical approach to variable selection. Next, we illustrate why transporting causal contrasts (as opposed to full counterfactual outcome distributions) may require fewer variables than a standard selection diagram may indicate. Finally, we categorize the different types of variables (according to causal structure) that might be included in transport estimators and use Monte Carlo simulations to evaluate how inclusion or exclusion of different variable types affects bias, variance, and mean squared error of the parametric g-computation transport estimator.

NOTATION AND DEFINITIONS
- Source population—population you are transporting results *from* (i.e. study sample)
- Target population—population you are transporting results *to*
- $P(Y^z)$ —the distribution of counterfactual outcome $Y$ if exposure $Z$ is assigned value $z$.
- $\Phi$ —a causal quantity that is a function of the counterfactual outcome distribution. For example, a causal contrast such as: $E(Y^{Z=1} - Y^{Z=0})$.
- S – selection node used in selection diagrams indicating population membership where $s \in \{0,1\}$ and $S = 1$ indicates the source population and $S = 0$ indicates the target population. These nodes are not standard random variables, but instead indicate where the data generating mechanisms may differ between the two populations.

- **TS** – a transport set defined as the set of variables included in a transport estimator. That is, any set of variables included in a transport estimator.
- **$TS_s$** – an s-admissible transport set defined as a transport set that d-separates all selection nodes (S, see above) from the outcome variable (such that $Y \perp S \mid TS_s$). There may be more than one s-admissible set for a given graph.
- **MSTS** – a minimally sufficient transport set. The smallest possible s-admissible transport set. There may be more than one **MSTS** for a given question.

TRANSPORTABILITY

    The goal of transportability is to determine what the results of a study conducted in one population (the *source population*) would have been had the study been conducted in a different population (the *target population*).[5] The intuition for why study results might differ depending on which population the study is conducted in is straightforward. If there are any characteristics that modify the effectiveness of the intervention and the distribution of those characteristics differs between the source and target populations, then we would expect that the intervention's effectiveness would similarly vary between the two populations.[6,7] If we can measure and account for those characteristics that both a) modify the effectiveness of the intervention and b) differ between two populations, we should be able to apply study results gathered in the source population to an external target population without having to repeat the entire study.

    The transportability framework formalizes this intuition and sets forth formal mathematical rules and conditions under which the results of a study can be transported from a source population to a target population.[1] We define $P(Y^z)$ as the counterfactual distribution of outcome $Y$ if exposure $Z$ is assigned value $z$ for all possible values of $Y$ and $Z$. This quantity can be thought of as the most general definition of a causal effect, as any causal contrast is a function of this counterfactual distribution. $P(Y^z)$ can be transported from a source population ($S = 1$) to a target population ($S = 0$) if the following assumptions are met:

1) S-admissibility (or conditional transport-exchangeability): $Y \perp S \mid TS_s$ where $TS_s$ is an s-admissible set. Note that $TS_s$ can be empty.
2) Transport-positivity:
$$\min_{z \in Z} P(S = 1, Z = z \mid TS_s = ts_s) > 0$$
for every $ts_s$ that has a positive density in the target population. That is, all values of the s-admissible set that are represented in the target population must have a non-zero probability of being in the source population and receiving exposure $Z = z$. Transport-positivity is met by default if TSs is empty.[2]

*Selection Diagrams for Variable Selection*

To illustrate the transportability framework in action, we use a toy example loosely motivated by the Finnish Geriatric Intervention Study to Prevent Cognitive Impairment and Disability (FINGER)[8]. Consider a hypothetical randomized controlled trial evaluating whether a multicomponent behavioral intervention was effective in reducing the 2-year risk of cognitive decline compared to standard of care among participants in Finland. Suppose the study found that randomization to a multicomponent behavioral intervention was effective in reducing the 2-year risk of cognitive decline, but we want to know what the results of this trial would have been had it been conducted in a US-based target population.

**Figure 1a** represents the true data generating mechanism for this toy example. For simplicity and without loss of generality, we assume that there are only two additional variables that might affect cognitive decline: systolic blood pressure >140 mmHg; and being a carrier of the apolipoprotein E-$\epsilon$4 (APOE-$\epsilon$4) variant. We define our outcome as risk of detectable cognitive decline during the 2 year follow-up period (detectable cognitive decline is defined as at least a 10% reduction in neurocognitive test battery score). We define our source population $S = 1$ as the Finnish study population and our target population $S = 0$ as the US-based target population; $X$ is randomized treatment assignment; $B = 1$ (systolic blood pressure > 140mmHg) and $G = 1$ (APOE-$\epsilon$4 carrier) both of which affect the risk of cognitive decline. $B$ and $G$ differ in distribution between the study and target populations.

Akin to how directed acyclic graphs (DAGs)[9,10] are used to select variables to control for confounding, selection diagrams are causal graphs used to determine which variables satisfy the s-admissibility criteria for transportability.[5] To create a selection diagram, we begin by drawing a DAG that represents the data-generating model for the source population. Next, we add selection nodes that indicate where there might be differences in the data-generating models between the source and target populations (**Figure 1b**). Selection nodes are not standard random variables; instead, they are indicators that point to the portions of the data-generating model that might differ between the two populations.[5,11]

Any set of variables that d-separates[10] all of the selection nodes from the outcome is an s-admissible set ($\boldsymbol{TS_s}$) for transporting $P(Y^Z)$ from S=1 to S=0. Note that a given graph may reveal more than one s-admissible set. Based on the selection diagram given in **Figure 1b**, the s-admissible set for this example is systolic blood pressure >140mmHg and APOE-$\epsilon$4 ($\boldsymbol{TS_s} = \{B, G\}$).

MINIMALLY SUFFICIENT TRANSPORT SET

A minimally sufficient transport set ($MSTS$) is the smallest possible s-admissible set that would satisfy assumption 1 for transporting a particular causal quantity from a source population to a target population. Though selection diagrams are useful for identifying s-admissible sets, in practice, they may not be able to isolate the $MSTS$ for two key reasons: 1) In many applications, we are interested in transporting a specific causal contrast (e.g., a risk difference) rather than a full counterfactual outcome distribution and 2) substantial uncertainty about the true data-generating model or how populations differ.

*Transportability of causal contrasts*

The transportability framework gives the assumptions and criteria for transporting the full counterfactual distribution of outcomes $P(Y^z)$ from the source population to the target population. However, in many applications, researchers may only be interested in transporting a particular causal quantity (e.g. a causal contrast or mean outcome value). If the causal quantity of interest ($\Phi$) is a function of $P(Y^z)$, then any set of variables that is s-admissible for transporting $P(Y^z)$ would also be s-admissible for transporting $\Phi$. However, there may be some variables that are necessary to transport $P(Y^z)$ that would be unnecessary for transporting $\Phi$.

For example, according to the selection diagram given in **Figure 1b**, the s-admissible set to transport $P(Y^z)$ includes both $B$ and $G$. This is also apparent from the structural equations in **Figure 1a**: $P(Y = 1)$ depends on both $B$ and $G$. However, suppose we are only interested in transporting the causal risk difference between those assigned to the intervention arm and those assigned to the treatment arm:
$$\Phi = P(Y^{Z=1} = 1) - P(Y^{Z=0} = 1)$$
$$\widehat{\Phi} = P(Y = 1|Z = 1) - P(Y = 1|Z = 0)$$
From the structural equations in **Figure 1a**, we see that this quantity only depends on $B$:
$$P(Y = 1|Z = 1) - P(Y = 1|Z = 0) = -.4B - .001(1 - B)$$
We can modify the selection diagram to reflect that we only want to transport this risk difference (represented as $\Phi$ in **Figure 1c**). The resulting graph indicates that the risk difference does not depend on $G$; only $B$ is required to d-separate the risk difference from the selection nodes.

The transport formula for transporting $P(Y^z)$ from the source to the target population using the transport set $\{B, G\}$ is:
$$P(Y^z|S = 0) = \sum_g \sum_b P(Y = 1|Z, G, B, S = 1)P(B, G|S = 0)$$

The transport formula for transporting $\Phi$ using the transport set $\{B\}$ is:

$$\Phi = P(Y^{Z=1} = 1 | S = 0) - P(Y^{Z=0} = 1 | S = 0)$$
$$= \sum_b P(Y = 1 | Z = 1, B, S = 1) P(B | S = 0)$$
$$- \sum_b P(Y = 1 | Z = 0, B, S = 1) P(B | S = 0)$$

**Table 1** shows the results of using each transport set (either $\{B, G\}$ or just $\{B\}$) to transport the mean outcome in each arm; the risk difference between arms; and the risk ratio between arms. We see that the transport set $\{B, G\}$, which includes both causes of the outcome, allows us to accurately transport all 3 quantities. Transport set $\{B\}$ is sufficient to transport our causal quantity of interest (the risk difference), but it is not sufficient to transport the risk ratio or average risk in each group.

This toy example illustrates that transporting a specific causal quantity may require fewer variables than would be necessary for transporting the full counterfactual distribution. In practice—when the true data-generating model is unknown—knowing which variables from the s-admissible set for transporting $P(Y^Z)$ are unnecessary for transporting $\Phi$ requires imposing parametric assumptions on the outcome-generating function that may be difficult to justify. As a result, researchers may reasonably choose to use the s-admissible set for $P(Y^Z)$ to avoid making these types of parametric assumptions at the expense of including potentially unnecessary variables in the transport estimators.

*Uncertainty in causal diagrams*

As with all causal graphs, excluding edges or selection nodes from a selection diagram is a stronger assumption than including them.[9] Given the uncertainty in many applied settings about the true data-generating model and how two populations might differ from one another, s-admissible sets derived using available knowledge will often include many more variables than would be necessary with perfect knowledge of the underlying data-generating model. How these extraneous variables might affect the performance of transport estimators is unclear.

SIMULATION EXPERIMENT

We conducted a Monte Carlo simulation experiment to examine the variable selection problem in transport estimators. Specifically, we explored how the inclusion of 5 different types of *unnecessary* variables (in addition to the MSTS) affect the bias, mean square error (MSE), and confidence interval coverage of the parametric g-formula transport estimator.[2] We limit our experiment to only consider variables that are not on the causal path from the exposure to the outcome.

*Classification of transport variables according to causal structure*

**Figure 2** shows 5 different variable types that might be unnecessarily included in an s-admissible set ($TS_s$). In this example, all variable types are subsets of the s-admissible set and are (by definition) not part of the $MSTS$; all types are mutually exclusive. The unnecessary variables are categorized according to their relationships to the selection nodes, outcome variable, and causal quantity conditional on a specific $MSTS$. Note that if a variable is not a cause of the outcome, it cannot be a cause of $\Phi$. If a variable is a cause of $\Phi$ it must also be a cause of the outcome.

*Simulation Experiments*

We generated data according to the following data-generating processes.

$$S \sim Ber(0.5)$$
$$Z \sim Ber(0.5)$$
$$MSTS, W_a, W_b \sim N(1 + 3S, sd_m)$$
$$W_c, W_d \sim N(1, 1)$$
$$W_e \sim N(0, 1)$$
$$Y \sim N(100 + 20Z + 10(MSTS)Z + 10(W_a) + 10(W_c)Z + 10(W_d), 5)$$

Where for each data-generating model $M = m$:
$$sd_m = \begin{cases} 1 + 5S, & M = 1 \\ 1 + 3S, & M = 2 \\ 1 + S, & M = 3 \end{cases}$$

The magnitude and likelihood of practical positivity violations[12] (also called random positivity[13]) was highest in data-generating model 3.

We aim to transport the causal quantity $\Phi = E(Y^{Z=1}) - E(Y^{Z=0})$ from the study population ($S = 1$) to the target population ($S = 0$). In all 3 data-generating models, the true value of $\Phi$ in the target population is 40 and the true value of $\Phi$ in the study population is 70.

For each data-generating model, we simulated 5000 datasets with a total N=5000 (with approximately 50% in $S = 1$ and 50% in $S = 0$). For each dataset, we fit a parametric g-computation transport estimator[2] for each of the following transport adjustment sets:

| Transport Adjustment Set ($TS_i$) |
| --- |
| $TS_1 = \{\text{MSTS}\}$ |
| $TS_2 = \{\text{MSTS}, W_a\}$ |
| $TS_3 = \{\text{MSTS}, W_b\}$ |
| $TS_4 = \{\text{MSTS}, W_c\}$ |
| $TS_5 = \{\text{MSTS}, W_d\}$ |
| $TS_6 = \{\text{MSTS}, W_e\}$ |
| $TS_7 = \{\text{MSTS}, W_a, W_b\}$ |
| $TS_8 = \{\text{MSTS}, W_a, W_c, W_d\}$ |
| $TS_9 = \{\text{MSTS}, W_c, W_d\}$ |
| $TS_{10} = \{\text{MSTS}, W_a, W_b, W_c, W_d, W_e\}$ |
| $TS_{11} = \{W_c\}$ |

**Table 2.** List of transport adjustment sets used with the parametric g-formula transport estimator for each simulation. $TS_7$ includes any variables that differ between the two populations; $TS_8$ includes all causes of Y; $TS_9$ includes the $MSTS$ and all causes of Y that don't differ between the two populations; $TS_{10}$ includes the full set of variables; and $TS_{11}$ does not meet the s-admissibility criterion and serves as a negative control.

To fit the g-formula transport estimators, we first fit a conditional linear regression in the source population ($S = 1$) regressing Y on Z and all variables in the transport set including all possible interaction terms. We then used this model to predict the values of Y in the target population setting $Z = 1$ and $Z = 0$ and calculated the difference in mean outcomes under each treatment assignment.[2] We used a non-parametric bootstrap with 1000 bootstrap samples to estimate the standard error.[14]

We report the estimated bias, variance, MSE, and confidence interval coverage for each transport set. All analyses were conducted using R version 3.5.2.[15]

*Simulation Results*

Across all 3 data-generating models, all transport sets that included the $MSTS$ (and therefore met the s-admissibility criterion) were unbiased (**Table 3**). However, using the $MSTS$ alone was not the optimal transport set in terms of MSE; $TS_8$ had the lowest MSE across all data-generating models. Among the s-admissible sets (all except $TS_{11}$), $TS_3$ had the highest MSE in each of the 3 data-generating models.

Excluding variables that were causes of the outcome but did not modify the causal quantity of interest ($W_a$ and $W_d$) did not negatively affect the bias of the estimators, and including unnecessary variables that were not causes of the outcome and that did not differ between the populations ($W_e$) did not increase the MSE compared to the MSTS alone.

Because of the smaller standard deviations for $MSTS$, $W_a$, and $W_b$ in data-generating model 3, this model was most likely to produce practical positivity violations. However, the parametric models in the estimators were correctly specified and could therefore accurately extrapolate beyond the bounds of the source data, so these practical positivity violations did not induce bias in the transport estimators.[12] Additionally, because the standard errors were smaller in the source population in this data-generating model compared to models 1 and 2, the estimates were generally more precise. However, the pattern of relative performance between the transport sets differed in this data generating model. $TS_8$, which included all causes of Y but no other unnecessary variables, performed substantially better than the other transport sets, while $TS_3$ and $TS_7$ had markedly higher MSEs.

DISCUSSION

Though selection diagrams play a vital role in identifying s-admissible sets for transporting a causal effect, in applied settings it is likely that these s-admissible sets may contain extraneous variables that are unnecessary for transporting the quantity of interest. The impact of including these extraneous variables on the parametric g-formula transport estimator's performance varies depending on the type of extraneous variable that is included.

In general, we observe that including all variables that are associated with the outcome—regardless of whether the variable modifies the causal quantity of interest or varies between the populations—improves the MSE by reducing the standard error. However, including variables that differ between the populations but are not associated with the outcome tends to increase the MSE.

There are several practical implications uncovered by this study. When faced with a variable selection problem for transport, it's best to focus on including as many causes of the outcome as possible. This is perhaps counterintuitive. Obvious differences between source and target populations may be the impetus for applying transportability methods in the first place, and these types of differences might be easier to detect. However, the strategy of including all variables that differ between two populations increases the chance of including extraneous variables that are not associated with the outcome, which would increase the MSE of the estimator.

Because we intended to highlight the impact of including different types of extraneous variables for the most common types of transport questions researchers are likely to face, we restricted our discussion and experiment to only include data-generating models with selection nodes on variables that were not affected by the treatment. Transporting results in situations where there are selection nodes directed at mediating variables requires additional measurements and assumptions that are beyond the scope of this manuscript. For background on transporting causal effects when selection nodes are directed at mediating variables, we point readers to Appendix 3 of Pearl and Bareinboim, 2011[5] and Bareinboim and Pearl, 2012[16] for more details.

Here, we only evaluated the parametric g-formula transport estimator. Other commonly used transport estimators include the inverse odds of selection weights[17] and doubly robust targeted maximum likelihood transport estimators.[18] We anticipate similar patterns across the different estimators, but future work should explore variable selection under other data-generating conditions and with other estimation approaches. Additionally, our simulation experiment only considered correctly specified parametric models in the g-formula transport estimators. As a result, the estimates were unbiased in spite of the practical positivity violations in data-generating model 3. If the models used in the estimators are not correctly specified, there is no guarantee that the estimates would be unbiased. Evaluation of transport estimators under incorrect model misspecification is an important topic for future work.

Based on our findings, a potential practical approach to variable selection for transportability using the g-formula estimator would be to use the study data alone to determine what variables should be measured in target populations to transport the results. For example, after completion of a trial, researchers could conduct a careful analysis to identify all the characteristics that modified the effect of interest. Researchers looking to transport the trial's results to a specific target population would then know what characteristics they would need to measure to do so. So long as the study measured all effect modifiers, this approach would ensure that the s-admissibility criteria is met and that any unnecessary variables included in the transport set would improve the precision of the estimates because they would all be causes of the outcome. Of course, trial results can only be transported if they enroll populations that are heterogeneous with respect to the effect modifiers and if all those effect modifiers are measured. Future work should explore data-driven approaches for identifying optimal transport sets to further improve the accuracy and precision of transport estimators.

|  |  | P(Y=1) | Risk Difference | Risk Ratio |
|---|---|---|---|---|
| Truth | $P(Y=1\|Z=0, S=0)$ | 0.680 | -0.121 | 0.823 |
|  | $P(Y=1\|Z=1, S=0)$ | 0.559 |  |  |
| Transported using $\{B, G\}$ | Transported $P(Y=1\|Z=0, S=0)$ | 0.680 | -0.121 | 0.823 |
|  | Transported $P(Y=1\|Z=1, S=0)$ | 0.559 |  |  |
| Transported using $\{B\}$ | Transported $P(Y=1\|Z=0, S=0)$ | 0.632 | -0.121 | 0.809 |
|  | Transported $P(Y=1\|Z=1, S=0)$ | 0.511 |  |  |

**Table 1** shows the transported risk difference and risk ratio adjusting for APOE-$\epsilon 4$ (G) and systolic blood pressure (B) or systolic blood pressure alone. If the target parameter is the risk difference, we see that adjusting for systolic blood pressure alone is sufficient.

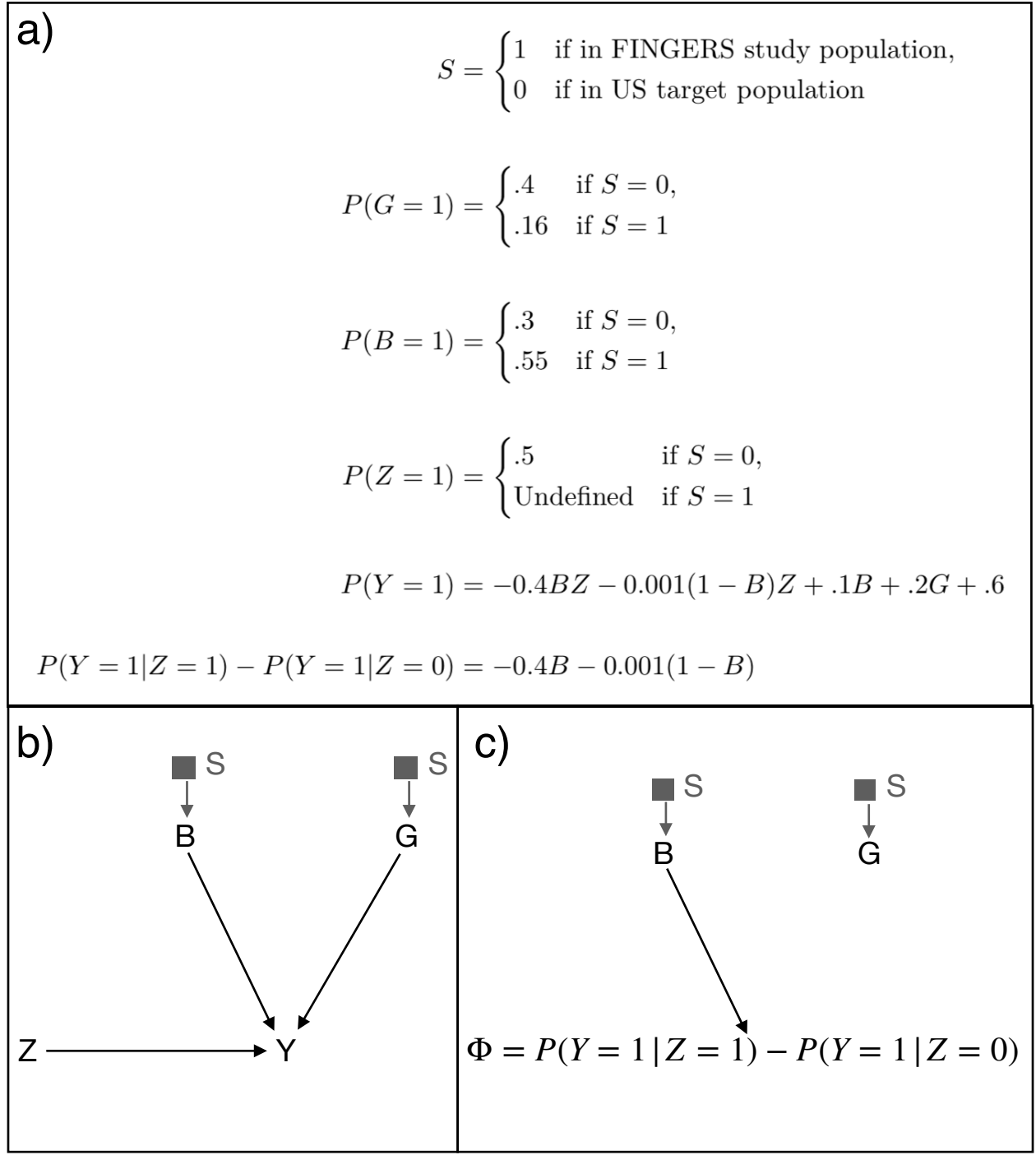

**Figure 1.** Structural causal model and corresponding selection diagrams for a toy example illustrating that fewer variables might be needed for transporting a causal contrast compared to transporting a full counterfactual outcome distribution. **Fig.1a** shows the true data-generating process for this example: $S$ indicates population; $G = 1$ indicates carrying at least one APOE- variant; $B = 1$ indicates systolic blood pressure >140mmHg; $Z$ indicates study arm; $Y = 1$ indicates having detectable cognitive decline by year 2. **Fig.1b** is the standard selection diagram based on this model. **Fig.1c** is the modified selection diagram focusing on the causal contrast of interest $\Phi$, the causal risk difference.

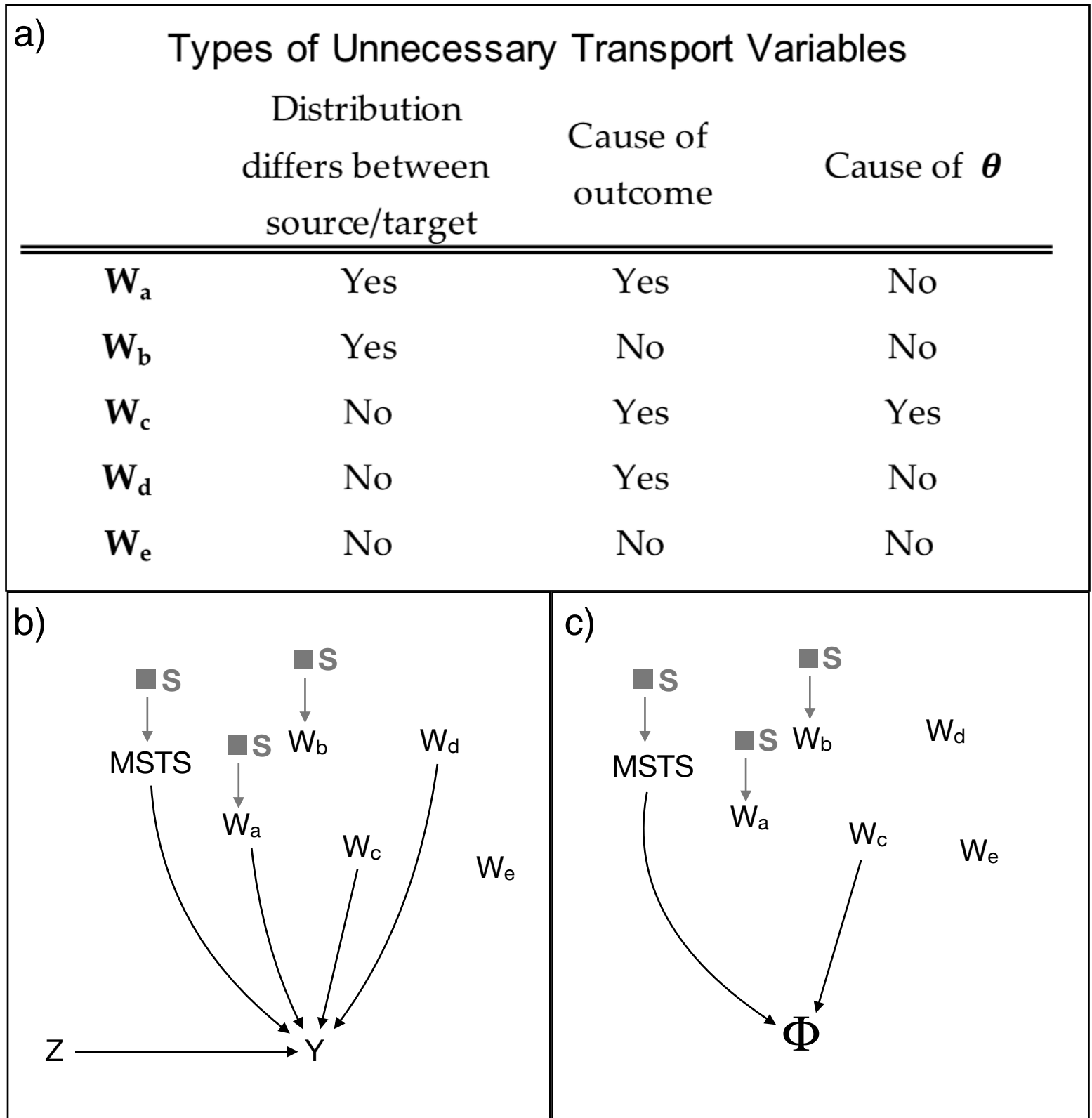

**Figure 2.** 5 categories of unnecessary transport variables. **Figure 2b-2c.** Selection diagrams showing the 5 categories of unnecessary variables that may be included in an s-admissible set. $MSTS$ is the minimally sufficient transport set. After conditioning on this set, $W_a$, $W_b$, $W_c$, $W_d$, and $W_e$ are all unnecessary to transport $\Phi$ across the populations.